\documentclass[aip,reprint,superscriptaddress]{revtex4-2}
\usepackage{graphicx}
\usepackage{dsfont}
\usepackage{amsmath}
\usepackage{amssymb}
\usepackage{bm}
\usepackage{physics}
\usepackage{subfigure}
\usepackage[colorlinks=true,linkcolor=blue,urlcolor=blue,citecolor=blue]{hyperref}
\usepackage{times}
\usepackage{soul}
\usepackage{changes}

\definechangesauthor[name=RF,color=orange]{RF}

\begin{document}

\newcommand{\Hop}{\hat{H}}
\newcommand{\Himp}{\hat{H}_{\rm imp}}
\newcommand{\gHimp}{\mathcal{H}_{\rm imp}}
\newcommand{\Hbath}{\hat{H}_{\rm bath}}
\newcommand{\Hhyb}{\hat{H}_{\rm hyb}}

\newcommand{\aop}{\hat{a}}
\newcommand{\adop}{\hat{a}^{\dagger}}
\newcommand{\sgp}{\hat{\sigma}^+}
\newcommand{\sgx}{\hat{\sigma}^x}
\newcommand{\sgy}{\hat{\sigma}^y}
\newcommand{\sgz}{\hat{\sigma}^z}
\newcommand{\nop}{\hat{n}}

\newcommand{\cop}{\hat{c}}
\newcommand{\cdop}{\hat{c}^{\dagger}}
\newcommand{\hc}{{\rm H.c.}}
\newcommand{\rhoop}{\hat{\rho}}
\newcommand{\rhoimp}{\hat{\rho}_{\mathrm{imp}}}
\newcommand{\rhobath}{\hat{\rho}_{\mathrm{bath}}}
\newcommand{\Uimp}{\hat{U}_{\mathrm{imp}}}
\newcommand{\Us}{\hat{U}_{\mathrm{S}}}
\newcommand{\Zimp}{Z_{{\rm imp}}}
\newcommand{\measure}{\mathcal{D}}
\newcommand{\gZ}{\mathcal{Z}}
\newcommand{\gK}{\mathcal{K}}
\newcommand{\gI}{\mathcal{I}}
\newcommand{\bolda}{\bm{a}}
\newcommand{\boldabar}{\bar{\bm{a}}}
\newcommand{\abar}{\bar{a}}
\newcommand{\im}{{\rm i}}
\newcommand{\contour}{\mathcal{C}}
\newcommand{\gA}{\mathcal{A}}
\newcommand{\gB}{\mathcal{B}}
\newcommand{\gM}{\mathcal{M}}
\newcommand{\gV}{\mathcal{V}}
\newcommand{\boldeta}{\bm{\eta}}
\newcommand{\boldetabar}{\bar{\bm{\eta}}}
\newcommand{\etabar}{\bar{\eta}}
\newcommand{\parity}{\mathcal{P}}
\newcommand{\current}{\mathcal{J}}

\title{Grassmann Time-Evolving Matrix Product Operators: an Efficient Numerical Approach for Fermionic Path Integral Simulations}
\author{Xiansong Xu}
\affiliation{College of Physics and Electronic Engineering, and Center for Computational Sciences, Sichuan Normal University, Chengdu 610068, China}
\affiliation{Science, Math and Technology Cluster, Singapore University of Technology and Design, 8 Somapah Road, Singapore 487372}
\author{Chu Guo}
\affiliation{Key Laboratory of Low-Dimensional Quantum Structures and Quantum Control of Ministry of Education, Department of Physics and Synergetic Innovation Center for Quantum Effects and Applications, Hunan Normal University, Changsha 410081, China}
\author{Ruofan Chen}
\email{physcrf@sicnu.edu.cn}
\affiliation{College of Physics and Electronic Engineering, and Center for Computational Sciences, Sichuan Normal University, Chengdu 610068, China}

\date{\today}
\begin{abstract}
  Developing numerical exact solvers for open quantum systems is a challenging task due to the non-perturbative and non-Markovian nature when coupling to structured environments. The Feynman-Vernon influence functional approach is a powerful analytical tool to study the dynamics of open quantum systems. Numerical treatments of the influence functional including the quasi-adiabatic propagator technique and the tensor-network-based time-evolving matrix product operator method, have proven to be efficient in studying open quantum systems with bosonic environments. However, the numerical implementation of the fermionic path integral suffers from the Grassmann algebra involved. In this work, we present a detailed introduction of the Grassmann time-evolving matrix product operator method for fermionic open quantum systems. In particular, we introduce the concepts of Grassmann tensor, signed matrix product operator, and Grassmann matrix product state to handle the Grassmann path integral.
  Using the single-orbital Anderson impurity model as an example, we review the numerical benchmarks for structured fermionic environments for real-time nonequilibrium dynamics, real-time and imaginary-time equilibration dynamics, and its application as an impurity solver. These benchmarks show that our method is a robust and promising numerical approach to study strong coupling physics and non-Markovian dynamics. It can also serve as an alternative impurity solver to study strongly-correlated quantum matter with dynamical mean-field theory.
\end{abstract}
\maketitle

\section{Introduction}

Isolated quantum systems are usually an idealization, as real-world quantum systems usually interact with external environments by design or unintentionally \cite{BreuerPetruccione2007,deVegaAlonso2017,WeimerOrus2021}. For example, even on the most sophisticated quantum computing devices to date, one has to inevitably take into account the dephasing noise \cite{Unruh1995, PalmaEkert1997}, which is a typical scenario of open quantum systems. The study of open quantum system dynamics is thus important to understand the behavior of these noisy quantum systems and to design corresponding protocols to mitigate the noise. In another scenario, quantum systems are intentionally coupled to environments that are modeled as collections of an infinite number of particles such as phonons or electrons to study transport properties \cite{caldeira1983-path,caldeira1983-quantum,Datta1995, HaugJauho2008,Ryndyk2016,LandiSchaller2022}. Such a scenario becomes the very foundational model for quantum transport, with a wide range of applications in developing quantum devices. Besides these direct modelings of realistic systems, open quantum systems can also be used to effectively describe the bulk systems in the application of dynamical mean-field theory (DMFT) \cite{georges1996-dynamical}. Lastly, we emphasize that open quantum systems are crucial in the study of the various fundamental aspects of statistical mechanics such as thermalization and equilibration \cite{PopescuWinter2006, GoldsteinZanghi2006,DAlessioRigol2016}, non-Markovian dynamics \cite{BreuerVacchini2016,LiWiseman2018, PollockModi2018,deVegaAlonso2017}, dissipative quantum chaos \cite{DenisovZyczkowski2019,AkemannProsen2019a,SaProsen2020a,HamazakiUeda2020}, and quantum thermodynamics \cite{GemmerMahler2004,BinderAdesso2018,DeffnerCampbell2019a,TalknerHanggi2020}.

Solving the dynamics of open quantum systems presents significant challenges in terms of both spatial complexity and temporal complexity. The problem of spatial complexity is twofold: i) the system part suffers from the exponential growth of the Hilbert space size; ii) methods that explicitly store the states of the environments would require additional spatial complexity. Temporal complexity is associated with the non-Markovian nature of open system dynamics with structured environments. An exact treatment of the open system dynamics formally requires all the past information of the system, which is generally impractical to obtain. Techniques developed focus on circumventing the above issues by imposing reasonable assumptions under different physical conditions. This gives rise to diverse techniques, each with its own flavor and different applicable regimes. 

A conceptually straightforward method is the family of tensor-network methods that evolve both the system and environments under the Schr\"odinger equation \cite{PriorPlenio2010,ChinPlenio2010a,TamascelliPlenio2019,LacroixChin2024}. The continuous environments are discretized into so-called star configurations and then mapped to semi-infinite chains. Further treatments can be implemented by imposing thermofield transformation to improve the efficiency for finite temperature simulations \cite{deVegaBanuls2015,GuoPoletti2018, XuPoletti2019a,KohnSantoro2021,KohnSantoro2022}. The path-integral formulation gives rise to various approaches including the hierarchical equation of motion techniques \cite{TanimuraKubo1989,JinYan2008a,LiYan2012a,DanShi2023,HuangChen2023,LambertNori2023,ZhangYan2024}, quasi-adiabatic propagator path integral (QuAPI) techniques \cite{makarov1994-path,makri1995-numerical,dattani2012-analytic}, continuous-time quantum Monte Carlo (CTQMC) \cite{gull2011-continuous, ErpenbeckCohen2023}, and non-pertubrative quantum master equations \cite{HuZhang1992,ZhangNori2012}. 
The nonequilibrium Green's function technique is another family of methods which derives from the propagator formalism \cite{SchwingerSchwinger1961,HaugJauho2008,WangLu2008,StefanuccivanLeeuwen2013,wang2013-nonequilibrium,Ryndyk2016}. Quantum master equation approaches are a huge family of methods that can be proposed phenomenologically, or derived in a perturbative manner \cite{Redfield1957,FlemingCummings2011,ThingnaHanggi2013,XuWang2017,HartmannStrunz2020}, and non-perturbatively from path integral methods \cite{HuZhang1992,ZhangNori2012} or formally from projection operator techniques \cite{Nakajima1958,Zwanzig1960}. 
For perturbative master equations, a mainstream approach is to kept the system-environment coupling at second order, which gives rise to the Redfield master equation \cite{Redfield1957}. 
Additional treatments can also be applied, including the reaction coordinates method \cite{HughesBurghardt2009,HughesBurghardt2009a,MartinazzoBurghardt2011,Iles-SmithNazir2016a,Anto-SztrikacsSegal2021}, canonical consistent master equation \cite{BeckerThingna2022a}, and the coarse graining approach \cite{SchallerBrandes2008,SchallerBrandes2009}. A family of master equation methods, including the pseudomode approach \cite{Garraway1997,TamascelliPlenio2018,LambertNori2019a,ChenGalperin2019} and the auxiliary master equation \cite{ArrigonivonderLinden2013a,DordaArrigoni2014a,SchwarzvonDelft2016}, is based on the concept of Markovian embedding by studying an extended effective system that captures non-Markovian dynamics under Markovian master equations.
Conditions for Markovian master equations known as the Gorini–Kossakowski–Sudarshan–Lindblad form are also given in Refs. \onlinecite{GoriniSudarshan1976,Lindblad1976}. We note that this brief introduction is far from complete. The pursuit of an efficient and accurate open quantum system solver has become even more critical in light of the rapid development of quantum devices, low-temperature electronics and their industrialization.     

Among all these methods, a particularly interesting numerical exact approach is the QuAPI method. Earlier studies focus on two-level systems with bosonic reservoirs \cite{makarov1995-stochastic,makarov1995-control,makri1997-universal,makri1997-stabilization,grifoni1998-driven}. Recently, it has been shown that QuAPI can be significantly enhanced with modern computational resources and tensor network techniques, which is known as the time-evolving matrix product operator (TEMPO) method \cite{strathearn2018-efficient}. The TEMPO method is now state-of-the-art for solving bosonic impurity problems. However, to the best of our knowledge, studying the fermionic counterpart is challenging due to the Grassmann algebra involved. Various other tensor-network-based influence functional methods have been proposed to circumvent the direct manipulation of Grassmann algebra \cite{ng2023-real,thoenniss2023-efficient,thoenniss2023-nonequilibrium,KlossAbanin2023,ParkChan2024}. Alternatively, we propose a Grassmann matrix product state which naturally handles Grassmann algebra \cite{chen2024-gtempo}, and develope the Grassmann TEMPO (GTEMPO) method which can be regarded as a fermionic version of the TEMPO method. The GTEMPO method allows us to construct a computational-friendly influence functional for fermionic systems. 

In this article, we first give a brief and formal review of the path integral formulations for bosonic and fermionic environments. We then present a detailed explanation of the GTEMPO method recently developed \cite{chen2024-gtempo,chen2024-egtempo,chen2024-real}, which consists of three main components: the vanilla QuAPI method, the Grassmann tensor and matrix product state, and the corresponding tensor-network treatment of the influence functional. A step-by-step introduction of the Grassmann tensor and matrix product state construction and associated arithmetic is given, followed by a detailed example on the single-orbital Anderson impurity model. We then review and summarize the numerical results, which benchmark against various other state-of-the-art methods for fermionic open quantum systems.

\section{Path integral formulations}

In this section, we present the general settings in the study of open quantum systems with $\hbar = k_{\rm B} = 1$.
The typical setup of an open quantum system includes the system part $\hat{H}_{\rm S}$ and the environment part $\hat{H}_{\rm E}$. They interact through the interaction term $\hat{H}_{\rm SE}$ and the total Hamiltonian is thus given by 
\begin{align}
  \hat{H} = \hat{H}_{\rm S} + \hat{H}_{\rm E} + \hat{H}_{\rm SE}.
\end{align}
The time evolution of the above composite system $\hat{H}$ is described by the Liouville--von Neumann equation 
\begin{align}
  i \frac{d \hat{\rho}(t)}{dt } = \qty[\hat{H}, \hat{\rho}(t)], 
\end{align}
where $\hat{\rho}(t)$ is the density matrix of the composite system. This equation yields a formal solution for the density matrix as follows
\begin{equation}
  \label{eq:rho-tot}
  \rhoop(t_f)=e^{-it_f\Hop}\rhoop(0)e^{it_f\Hop}.
\end{equation}

Other typical conditions include the decoupled initial condition where we assume there is no initial correlation between the system and the environment, i.e., 
\begin{align}
  \label{eq:separate}
  \hat{\rho}(0) = \hat{\rho}_{\rm S}(0) \otimes \hat{\rho}_{\rm E}.
\end{align}    
The environment is in Gibbs state defined by $\hat{\rho}_{\rm E} = e^{-\beta\Hop_{\rm E}}$, and is modeled as a collection of an infinite number of bosons or fermions, for which
\begin{align}
  \hat{H}_{\rm E} &= \sum_k \omega_k \hat{b}_k^\dagger \hat{b}_k,\ \text{bosons},\label{eq:bosonbath}\\
  \hat{H}_{\rm E} &= \sum_k \omega_k \hat{c}_k^\dagger \hat{c}_k,\ \text{fermions}. \label{eq:fermionbath}
\end{align}
The coupling between the environment modes and the system is characterized by the spectral density 
\begin{align}
  J(\omega)  = \sum_k V_k^2 \delta(\omega - \omega_k). 
\end{align}

In the seminal work by Feynman and Vernon \cite{feynman1963-the}, the quantum Brownian motion was modeled by a particle coupled to a bath of harmonic oscillators, using first quantization notation. For notational simplicity and to better correspond with the fermionic case, we consider the second quantized environment Hamiltonian in this section. We will formally derive the system partition function through the path integral formalism for both bosonic and fermionic cases. This helps to clarify and visualize the similarities and differences between the bosonic and fermionic path integrals.

\subsection{Bosonic environments}

We start with the case where the environment is a collection of an infinite number of bosons. The environment Hamiltonian is thus given by Eq. \eqref{eq:bosonbath} and the corresponding system-environment coupling is given by
\begin{align}
  \hat{H}_{\rm SE} = \hat{s} \sum_k V_k (\hat{b}^\dagger_k + \hat{b}_k)
\end{align}
where $\hat{s}$ is the system operator and is Hermitian. Since the total density matrix at the time $t_f$ is given by Eq.~\eqref{eq:rho-tot}, we can formally discretize the time evolution operator into $N$ pieces with $N\delta t  = t_f$ for which
\begin{align}
  \hat{\rho}(t_f) = e^{-i\delta t \hat{H}} \cdots e^{-i\delta t \hat{H}} \hat{\rho}(0) e^{i\delta t \hat{H}} \cdots e^{i\delta t \hat{H}}.
  \label{eq:rho-tot-discretize}
\end{align}
The partition function of the composite system is given by 
\begin{align}
  Z(t_f) =  \tr[\hat{\rho}(t_f)],
  \label{eq:partition-function}
\end{align}
and the system partition function is defined as
\begin{equation}
  Z_{\rm S}(t_f)=Z(t_f)/Z_{\rm E},\quad Z_{\rm E}=\tr e^{-\beta\Hop_{\rm E}},
\end{equation}
where $Z_{\rm E}$ is the partition function of the environment.

\begin{figure}
  \includegraphics{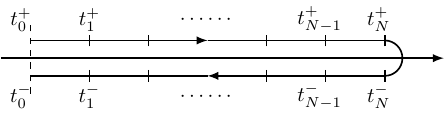}
  \caption{The Keldysh contour $\mathcal{C}$ for the discretized time points $t^+_0, t^+_{1}, \cdots, t_{N}^+, t_N^-, \cdots,t_{1}^-,t^-_0$. The upper is the forward branch and the lower is the backward branch.}
  \label{fig:keldysh}
\end{figure}

It is convenient to introduce coherent states to handle the degrees of freedom of the environment \cite{negele1998-quantum}. The bosonic coherent state $\ket{\varphi}$ is defined as the eigenstate of environment annihilation operator $\hat{b}_k$ for which
\begin{equation}
  \hat{b}_k \ket{\varphi} = \varphi_k \ket{\varphi},\quad\bra{\varphi} b_k^\dagger =\bra{\varphi} \bar{\varphi}_k,
\end{equation}
 where $\varphi_k$ is a complex number with $\bar{\varphi}_k $ being its complex conjugate. The identity operator for the composite system can then be represented in terms of the eigenbasis of the system operator $\hat{s}$ and the bosonic coherent state $\ket{\varphi}$  as
\begin{align}
  \sum_s \ket{s}\bra{s} 
  \int \left[\prod_k \frac{d \bar{\varphi}_k d \varphi_k}{2\pi i} e^{-\bar{\varphi}_k\varphi_k} \right] \ket{\varphi}\bra{\varphi} = 1.
  \label{eq:boson-closure}
\end{align}
We can now insert this identity operator between all the neighboring exponents in the trace, and then the corresponding partition function for the composite system is given by 
\begin{align}
  Z(t_f) &= \sum_{s^{\pm}_0,\cdots,s^{\pm}_N}  \int \qty[\prod^N_{\alpha=0} \prod_k \frac{d \bar{\varphi}^{\pm}_{\alpha k} d \varphi^{\pm}_{\alpha k}}{2\pi i} e^{-\bar{\varphi}^{\pm}_{\alpha k}\varphi^{\pm}_{\alpha k}}] \nonumber \\
  &\times \bra{s^+_N \varphi^+_N} e^{-i \delta t \hat{H}} \ket{s^+_{N-1}\varphi^+_{N-1}} \nonumber \\
  &\cdots \bra{s^+_0 \varphi^+_0} \hat{\rho}(0)\ket{s^-_0\varphi^-_0} \nonumber \\
  &\cdots \bra{s^-_{N-1} \varphi^-_{N-1}} e^{-i \delta t \hat{H}} \ket{s^-_N \varphi^-_N},
  \label{eq:boson-partition-total}
\end{align} 
where we denote $s(t^+_k),s(t^-_k)$ as $s^+_k,s^-_k$ and the boundary conditions are given by $s^{+}_N = s^-_N$ and $\varphi^{+}_N = \varphi^-_N$. Here, we have labeled the time points from left to right as $(t^+_N, t^+_{N-1}, \cdots t_0^+, t_0^-, \cdots, t^-_N)$ and these time points form a closed contour $\mathcal{C}$ known as the Keldysh contour as shown in Fig. \ref{fig:keldysh} \cite{keldysh1965-diagram,lifshitz1981-physical,kamenev2009-keldysh,wang2013-nonequilibrium}. 

After integrating out all the bath variables $\bar{\varphi}^{\pm}_{\alpha k},\varphi^{\pm}_{\alpha k}$, we can obtain the system partition function as
\begin{align}
  \label{eq:bosonic-path-integral}
  Z_{\rm S}(t_f)= \sum_{s^{\pm}_0,\cdots,s^{\pm}_N} \mathcal{K}[s^{\pm}_0,\cdots,s^{\pm}_N] \mathcal{I}[s^{\pm}_0,\cdots,s^{\pm}_N],
\end{align}
with the boundary condition $s^+_N = s^-_N$. In the continuous-time limit $\delta t \to 0$, the above partition function  can be written in the path integral form as
\begin{align}
  Z_{\rm S}(t_f) = \int\mathcal{D}[\bm{s}] \mathcal{K}[\bm{s}] \mathcal{I}[\bm{s}],
\end{align}
where we have simplified our notation by denoting $\bm{s} = (s^+_N, \cdots, s^+_0, s^-_0, \cdots, s^-_N)$. Here $\mathcal{D}[\bm{s}]$ is the measure and $\int\mathcal{D}[\bm{s}]$ stands for the summation in Eq. \eqref{eq:bosonic-path-integral}. Note that here the complex conjugate of $s_k^{\pm}$ is just itself as $\bar{s}_k^{\pm}=s_k^{\pm}$. The propagator $\mathcal{K}[\bm{s}]$ describes the free evolution of the system part and is given by
\begin{align}
  \gK[\bm{s}]=\lim_{\delta t\rightarrow 0}&\bra{s^+_N} e^{-i \delta t \hat{H}_{\rm S}} \ket{s^+_{N-1}} \cdots \bra{s^+_0} \hat{\rho}_{\rm S}(0)\ket{s^-_0} \nonumber \\&\times \cdots \bra{s^-_{N-1}} e^{i\delta t \hat{H}_{\rm S}} \ket{s^-_N}.
\end{align}
The influence functional $\gI[\bm{s}]$ contains the effect of the environment on the system and is given by
\begin{align}
  \gI[\bm{s}] = \exp\left[-\int_{\mathcal{C}} dt \int_{\mathcal{C}} dt' \bar{s}(t) \Delta(t,t') s(t')\right],
  \label{eq:boson-IF}
\end{align}
where the integrals are along the Keldysh contour $\mathcal{C}$. The correlation function $\Delta(t,t')$ is defined as
\begin{align}
  \Delta(t,t') = \int J(\omega) D_\omega(t,t') d\omega,
\end{align}
where $D_\omega(t,t')$ is the free environment contour-ordered Green's function defined as 
\begin{align}
  D_\omega(t,t') = \langle T_{\mathcal{C}} \hat{b}_{\omega}(t)\hat{b}_{\omega}^\dagger(t') \rangle_0 = \begin{cases}
    \langle \hat{b}_{\omega}(t)\hat{b}_{\omega}^\dagger(t') \rangle_0, & t \succeq   t' \\
    \langle \hat{b}_{\omega}^\dagger(t')\hat{b}_{\omega}(t) \rangle_0, & t \prec t'
  \end{cases}.
\end{align} 
We use $\succeq$ and $\prec$ to denote the order of $t$ and $t'$ on the Keldysh contour.  

We can denote $\gK[\bm{s}]\gI[\bm{s}]$ as $\gZ[\bm{s}]$, which is named the augmented density tensor (ADT). Formally, to evaluate the expectation value of the system operator $\hat{s}$ at some time $t$, we can insert the operator at the corresponding time point in the path integral expression of the composite system given by Eq.~\eqref{eq:boson-partition-total}.  Note that the operator can be inserted on either the forward or the backward branch of the Keldysh contour due to to the cyclic property of the trace.  For example, we can evaluate the expectation value $\expval*{\hat{s}(t)}$ as
\begin{equation}
 \expval*{\hat{s}(t)} 
 = \frac{1}{Z}\tr[  e^{-i \hat{H} (t_f - t)}  \hat{s} e^{-i \hat{H} t} \hat{\rho}(0)e^{i \hat{H} t_f }],
\end{equation}
which, in the path integral language, is given by 
\begin{equation}
  \expval*{\hat{s}(t)}=Z_{\rm S}^{-1}(t_f)\int\mathcal{D}[\bm{s}] \mathcal{Z}[\bm{s}] s_k^+,\ k\delta t = t.
\end{equation}
Equivalently, the expectation value $\expval{\hat{s}(t)}$ can be written as
\begin{equation}
  \expval{\hat{s}(t)} =\frac{1}{Z}\tr[  e^{-i \hat{H} t_f } \hat{\rho}(0) e^{i \hat{H} t} \hat{s} e^{i \hat{H} (t_f -t )}],
\end{equation}
where the operator $\hat{s}$ is now inserted on the backward branch instead of the forward branch. With the path integral formalism, we have 
\begin{equation}
  \expval{\hat{s}(t)}=Z_{\rm S}^{-1}(t_f)\int_{s(t_f^-)=s(t_f^+)} \mathcal{D}[\bm{s}] \mathcal{Z}[\bm{s}] s_k^-,\ k\delta t = t.
\end{equation}

Similarly, the two-time correlation $\expval*{\hat{s}(t)\hat{s}(t')}$ (assuming $t>t'$) can be found by inserting the system operator $\hat{s}$ to the corresponding time points $t$ and $t'$  on the Keldysh contour using the cyclic property of the trace. Depending on the branches where the operators are inserted, we have three combinations, i.e.,  
\begin{align}
  \langle \hat{s}(t) \hat{s}(t') \rangle =&  Z_{\rm S}^{-1}(t_f) \int_{s(t_f^-)=s(t_f^+)} \mathcal{D}[\bm{s}] \mathcal{Z}[\bm{s}] s^+_j s^+_k\\
  =&Z_{\rm S}^{-1}(t_f) \int_{s(t_f^-)=s(t_f^+)} \mathcal{D}[\bm{s}] \mathcal{Z}[\bm{s}] s^-_j s^+_k\\
    =&Z_{\rm S}^{-1}(t_f) \int_{s(t_f^-)=s(t_f^+)} \mathcal{D}[\bm{s}] \mathcal{Z}[\bm{s}] s^-_j s^-_k,
\end{align}
where $j\delta t = t, k\delta t = t'$.

\subsection{Fermionic environments}
The fermionic environment Hamiltonian is given by Eq. \eqref{eq:fermionbath}. We consider the system-environment coupling in the form of
\begin{align}
  H_{\rm SE} = \sum_k V_k (\hat{a}^\dagger \hat{c}_k + \hat{c}_k^\dagger \hat{a}),
\end{align}
where $\hat{a}$ ($\hat{a}^{\dag}$) is the fermionic annihilation (creation) operator for the system. This system-environment coupling term is also known as hybridization term. The corresponding fermionic coherent states for the system and environment are defined through  
\begin{align}
  \hat{a} \ket{a} = a \ket{a}, \quad \bra{a}\hat{a}^\dagger = \bra{a}\bar{a} , \\
  \hat{c}_k \ket{c} = c_k \ket{c}, \quad \bra{c}\hat{c}^\dagger_k = \bra{c} \bar{c}_k,
\end{align}
where $a$, $\bar{a}$, $c_k$, $\bar{c}_k$ are Grassmann variables (G-variables), and $\bar{a},\bar{c}_k$ are Grassmann conjugates of $a,c_k$ \cite{negele1998-quantum}.  The use of G-variables poses a challenge in the numerical evaluation of the fermionic path integral since they cannot be manipulated as the ordinary numbers. 

Now we try to formally evaluate the partition function. We repeat the procedures from Eqs. (\ref{eq:rho-tot-discretize}-\ref{eq:boson-partition-total}) except that the closure relation is now given by 
\begin{align}
  \int d\bar{a} da e^{-\bar{a} a}\int\left[ \prod_k d \bar{c}_k d c_k e^{-\bar{c}_k c_k}\right]\ketbra{ac}{ac} = 1,
\end{align}
where $\int$ stands for the Grassmann integral (G-integral). Then the total partition function can be written as
\begin{align}
  Z(t_f) & = 
  \int \prod^N_{\alpha=0} d\bar{a}_{\alpha} d a_{\alpha} e^{- \bar{a}_\alpha a_{\alpha}} \int \prod^N_{\alpha=0} \prod_k d\bar{c}_{\alpha k} d c_{\alpha k} e^{-\bar{c}_{\alpha k} c_{\alpha k}} \nonumber \\
  & \bra{(-a^+_N)(-c^+_N)} e^{-i \delta t \hat{H}} \ket{a^{+}_{N-1} c^{+}_{N-1}} \nonumber \\
  & \cdots   \bra{a^{+}_0 c^{+}_0} \rho(0)\ket{a^-_0c^-_0} \nonumber\\ 
  & \cdots \bra{a^-_{N-1} c^-_{N-1}} e^{-i \delta t \hat{H}} \ket{a^-_N c^-_N}.
\end{align}
Note that the coherent state $\bra{(-a_N^+)(-c_N^+)}$ on the leftmost side has extra minus signs according to fermionic boundary condition $a_N^+=-a_N^-,c_{Nk}^+=-c_{Nk}^-$. After integrating out the environment variables, the system partition function can be written in the path integral formalism as
\begin{align}
 Z_{\rm S}(t_f) = \int\mathcal{D}[\bar{\bm{a}} \bm{a}] \gK[\bar{\bm{a}}, \bm{a}] \gI[\bar{\bm{a}}, \bm{a}],
 \label{eq:fermion-system-partition}              
\end{align}
where the measure is given by
\begin{equation}
\mathcal{D}[\boldabar\bolda]=\prod_td\abar(t)da(t)e^{-\abar(t)a(t)}.
\end{equation}
Note that unlike the bosonic environment where $\bar{s}_k^{\pm} = s_k^{\pm}$, we have to keep track of both $\bar{a}_k^{\pm}$ and $a_k^{\pm}$ for fermions. The free propagator is given by
\begin{align}
  \gK[\bar{\bm{a}},\bm{a}] &= \bra{(-a^+_N)} e^{-i \delta t \hat{H}_{\rm S}} \ket{a^{+}_{N-1}} \nonumber \\ &\cdots   \bra{a^{+}_0} \hat{\rho}_{\rm S}(0)\ket{a^-_0} \cdots \bra{a^-_{N-1}} e^{-i \delta t \hat{H}_{\rm \rm S}} \ket{a^-_N}.
  \label{eq:fermion_K1}
\end{align}
The corresponding influence functional is given by
\begin{align}
  \gI\left[\bar{\bm{a}}, \bm{a}\right] = \exp\left[-\int_{\mathcal{C}} dt \int _{\mathcal{C}} dt' \bar{a}(t) \Delta(t,t') a(t')\right]
  \label{eq:fermion_IF}
\end{align}
where $\Delta(t,t')$ is usually known as the hybridization function and is given by
\begin{align}
  \Delta(t,t') = \mathcal{P}_{tt'} \int d \omega J(\omega) D_{\omega}(t,t'),
  \label{eq:hybridization}
\end{align}
where $D_{\omega}(t,t')$ is the free fermionic contour-ordered Green's function for the environment 
\begin{align}
  D_{\omega}(t,t') = \langle T_{\mathcal{C}} c_{\omega}(t)c_{\omega}^\dagger(t') \rangle_0 = \begin{cases}
    \langle c_{\omega}(t)c_{\omega}^\dagger(t') \rangle_0, & t \succeq   t' \\
    -\langle c_{\omega}^\dagger(t')c_{\omega}(t) \rangle_0, & t \prec t'
  \end{cases}.
\end{align} 
Here $\succeq$ and $\prec$ refer to the order of $t$ and $t'$ on the Keldysh contour, and $\mathcal{P}_{tt'}=1$ when $t,t'$ are on the same branch, otherwise $\mathcal{P}_{tt'}=-1$.

Similar to the bosonic case, we can define the ADT as $\mathcal{Z}\left[\bar{\bm{a}}, \bm{a}\right] = \gK[\bar{\bm{a}},\bm{a}] \gI[\bar{\bm{a}},\bm{a}]$. Formally, from the ADT we can measure any observables or correlation functions. For example, a two-time correlation function can be obtained (assuming $t>t'$) as
\begin{align}
 \langle \hat{a}(t)\hat{a}^\dagger(t') \rangle = Z^{-1}_{\rm S} (t_f) \int D[\bar{\bm{a}} \bm{a}] \mathcal{Z}[\bar{\bm{a}},\bm{a}] a^+_j \bar{a}^+_k
 \label{eq:fermionic_corr}
\end{align}
where $j \delta t = t, k \delta t = t'$.
Similar to the bosonic case, the operators can be inserted on different branches giving three possible combinations.

The above formulation for bosonic and fermionic path integral is formally complete but it does not give a practical way to evaluate the path integral numerically. 
In the following sections, we will briefly introduce the QuAPI and TEMPO methods to handle numerical evaluation of the bosonic path integrals, which set the basis of the GTEMPO method.

\section{From QuAPI to TEMPO}

In 1994, Makarov and Makri introduced the quasi-adiabatic propagator path integral techniques to handle the bosonic influence functional numerically \cite{makarov1994-path}. The QuAPI method has been extensively used to explore bosonic dissipative open systems such as dissipative driven quantum systems \cite{makarov1995-control,makarov1995-stochastic,dong2004-optimizing,dong2004-quantum,makri1997-stabilization,makri1997-universal,nalbach2009-landau,thorwart2000-iterative}, anharmonic environments \cite{ilk1994-real,makri1997-universal}, linear response \cite{makri1999-linear}, and also beyond Keldysh contour \cite{shao2002-iterative}.

The core idea is to discretize the path $s^{\pm}(t)$ into segments of equal length $(j-1/2)\delta t \le t \le (j+1/2) \delta t$. Within each segment, $s^\pm(t) = s^\pm_j$ remains unchanged. The discretized influence functional can be written as 
\begin{align}
  I(s^\pm_0, s^\pm_1, \cdots, s^\pm_N) = \exp\left(-\sum_{\zeta\zeta'}\sum^N_{j=0} \sum^N_{k=0} s^\zeta_j \Delta^{\zeta\zeta'}_{jk} s^{\zeta}_k\right)
\end{align}
where $\zeta,\zeta' = \pm$ and $\Delta_{jk}^{\zeta\zeta'}$ is the hybridization function after discretization. The QuAPI method also requires that the hybridization function decays significantly within influence functional to perform finite memory truncation. Such a truncation will enable an iterative scheme to evolve the reduced density matrix.

However, the QuAPI method can still be limited by the length of the memory kernel as the temporal complexity of the influence functional grows exponentially with the number $N$ of discretized timesteps. Since the influence functional is a high-rank tensor, tensor-network techniques could offer significant advantages in dimensional reduction \cite{schollwoeck2005-density,schollwoeck2011-density,orus2014-practical}. Strathearn et al. introduced the time-evolving matrix product operator method to represent the bosonic ADT as the matrix product state \cite{strathearn2018-efficient}. Note that the bosonic ADT can also be described using the language of the process tensor  \cite{CostaShrapnel2016,PollockModi2018,pollock2018-tomographically,joergensen2019-exploiting}. The key idea of TEMPO is to write the discretized influence functional $\gI{[\bm{s}]}$ as the product of the following partial influence functional (PIF) as
\begin{equation}
  \gI[\bm{s}]=\prod_{j,\zeta}\gI_j^{\zeta}[s]=\prod_{j,\zeta}e^{-s_j^{\zeta}\sum_{k,\zeta'}\Delta^{\zeta\zeta'}_{jk}s^{\zeta'}_k}.
\end{equation}
The above expression allows a matrix product state representation of each PIF, and then the IF can be constructed through the multiplication of the corresponding PIFs. Such a representation significantly enhances the capability of the vanilla QuAPI method in terms of computational efficiency and memory length storage. In fact, it can be even unnecessary to perform a memory length truncation as in the QuAPI method.

The TEMPO method is now considered state-of-the-art for studying bosonic open quantum systems. It has been applied to study various fundamental aspects of open quantum systems, including multi-time correlation \cite{joergensen2020-discrete}, equilibration and thermalization \cite{chiu2022-numerical, fux2023-thermalization}, environment dynamics \cite{gribben2021-using}, and nonadditive environment \cite{gribben2022-exact}. Other applications include the study of phase transition \cite{otterpohl2022-hidden}, optimal control \cite{fux2021-efficient}, quantum stochastic resonance \cite{chen2023-non}, and nonequilibrium heat current \cite{popovic2021-quantum,chen2023-heat}. The corresponding tensor network structures within TEMPO are also discussed in Refs. \onlinecite{joergensen2019-exploiting,ye2021-constructing,bose2022-pairwise,cygorek2024-sublinear,link2024-open}.

However, for fermionic systems, it is not straightforward to handle the influence functional numerically due to the Grassmann algebra involved. There are various attempts inspired by QuAPI including Refs. [\onlinecite{SegalReichman2010,SegalReichman2011,SimineSegal2013,chen2019-dissipative,chen2020-landau}] which integrate out the discretized environment through Blankenbecler-Scalapino-Sugar (BSS) identity \cite{blankenbecler1981-monte} and Levitov’s formula \cite{klich2003-quantum,abanin2004-tunable,abanin2005-fermi}. More recently, Ng et al. \cite{ng2023-real} and Thoenniss et al. \cite{thoenniss2023-efficient,thoenniss2023-nonequilibrium} convert the Grassmann path integral formalism back into the Fock state basis to avoid the direct manipulation of the G-variables. In the following sections, we will show how to construct Grassmann tensor (G-tensor) and the Grassmann matrix product states (GMPS) to achieve direct numerical evaluation within the Grassmann algebra.

\section{QuAPI for fermions}

Before we define the Grassmann matrix product state, we present the essential ingredients of the QuAPI method for fermionic systems. We discretize the IF following the same spirit as the bosonic case. On the normal time axis, the G-variables are split into two branches $\abar^{\pm}(t),a^{\pm}(t)$, and accordingly, the hybridization defined in Eq. \eqref{eq:hybridization} is split into four blocks
\begin{equation}
  \Delta(t,t')=\mqty[\Delta^{++}(t,t') & \Delta^{+-}(t,t')\\
  \Delta^{-+}(t,t') & \Delta^{--}(t,t')].
\end{equation}
We split $t_f=N\delta t$ and the trajectories
$\abar^{\pm}(t),a^{\pm}(t)$ into intervals of equal
duration as $(j-\frac{1}{2})\delta t<t<(j+\frac{1}{2})\delta t$, then
the hybridization function is discretized as
\begin{equation}
  \Delta^{\zeta\zeta'}_{jk}=\int_{(j-\frac{1}{2})\delta t}^{(j+\frac{1}{2})\delta t}\dd{t}
  \int_{(k-\frac{1}{2})}^{(k+\frac{1}{2})\delta t}\dd{t'}\Delta^{\zeta\zeta'}(t,t'),
\end{equation}
where $\zeta,\zeta'=\pm$ indicates the branch on the Keldysh contour, and $0\le j,k\le N$. This procedure leads to the discretized IF
\begin{equation}
  \gI[\boldabar,\bolda]=e^{-\sum_{\zeta\zeta'}\sum_{jk}\abar_{ j}\Delta^{\zeta\zeta'}_{jk}a_{j}}.
\end{equation}
In terms of G-variables, the discretized $\gK$ can be written as
\begin{equation}
  \begin{split}
    \gK[\boldabar,\bolda]=
    &\mel{-a_N}{\Us}{a_{N-1}^+}\cdots\mel{a_1^+}{\Us}{a_0^+}\\
    &\times\mel{a_0^+}{\hat{\rho}_{\rm S}(0)}{a_0^-}\\
    &\times\mel{a_0^-}{\Us^{\dag}}{a_1^-}\cdots\mel{a_{N-1}^-}{\Us^{\dag}}{a_N},
  \end{split}
\end{equation}
where $\Us=e^{-i \hat{H}_{\rm S} \delta t}$. It should be noted that here we remove the branch superscript $\pm$ of the boundary G-variables $\abar_N,a_N$, which indicates that they are connected according to the definition of the trace. Although such a setting of boundary G-variables is natural, it is inconvenient for the QuAPI scheme used: $\abar_N,a_N$ in fact do not belong to the same branch on the contour as $\abar_N$ is at $t_f^+$ and $a_N$ is at $t_f^-$. To resolve such an issue, we introduce extra G-variables $\abar,a$ to take care of the boundary conditions. Hence $\gK$ can be equivalently written as
\begin{equation}
  \label{eq:K}
    \begin{split}
    \gK[\boldabar,\bolda]=
    &\braket{-a}{a_N^+}\mel{a_N^+}{\Us}{a_{N-1}^+}\cdots\\
    &\times\cdots\mel{a_{N-1}^-}{\Us^{\dag}}{a_N^-}\braket{a_N^-}{a}.
  \end{split}
\end{equation}
Now at time $t_f^+$ there is a pair of G-variables $\abar_N^+,a_N^+$,
and the pair $\abar_N^-,a_N^-$ is at time $t_f^-$.

\section{Naive MPS representation for Grassmann Tensor}

Since the TEMPO method relies on the tensor network expression for the (bosonic) influence functional, it is thus natural to consider the fermionic tensor network to express the Grassmann influence functional. We highlight that the existing fermionic tensor-network methods \cite{FidkowskiKitaev2011,BultinckVerstraete2017,MortierVanderstraeten2024} and the high-dimensional Grassmann tensor networks \cite{GuGu2013,YoshimuraSakai2018,AkiyamaKadoh2021} share similar spirit as the main goal is to deal with the fermionic statistics and anti-commutation relations. 

To suit the expression of fermionic influence functional, we introduce the Grassmann tensor (G-tensor) to represent the discretized $\mathcal{K}$ and $\gI$. The
discretized $\gK$ and $\gI$ are essentially G-tensors spanned
by the G-variables $\{\boldabar,\bolda\}$. 

For an algebra of G-variables of $n$ components $\xi_1,\ldots,\xi_n$, we define the Grassmann tensor $\mathcal{A}$ with
components 
\begin{equation}
  \label{eq:g-tensor}
  \mathcal{A}^{i_1\cdots i_n}=A^{i_1\cdots i_n}\xi_1^{i_1}\cdots\xi_n^{i_n},\quad i_k=\{0,1\},
\end{equation}
where the superscript $i_k$ over the G-variables $\xi_k$ represents actual powers. A number in the Grassmann algebra is given by the contraction of
a certain G-tensor
$\sum_{i_1,\ldots,i_n}\mathcal{A}^{i_1\cdots i_n}$, and the coefficients
tensor $A^{i_1\cdots i_n}$ is a rank-$n$ array of $2^n$ scalars.  

Formally, the coefficient tensor $A^{i_1\cdots i_n}$ can be represented as an MPS $A^{i_1}\cdots A^{i_n}$ exactly, i.e., 
\begin{equation}
  A^{i_1\cdots i_n} \equiv  A^{i_1}\cdots A^{i_n}.
\end{equation}
Next, we will define the corresponding G-tensor multiplications which are used to construct $\gK,\gI$ and finally $\gZ$.

\subsection{Multiplying G-variables to a G-tensor}

Given a G-tensor $\mathcal{A}$ generated by $\xi_1,\ldots,\xi_n$, 
we can obtain a new G-tensor $\mathcal{B}$ by multiplying a single G-variable to $\mathcal{A}$ 
\begin{align}
\mathcal{B}=\xi_k\mathcal{A}.
\end{align}
In the MPS representation, this is denoted as 
\begin{align}
\mathcal{B}=\xi_k A^{i_1}\cdots A^{i_n} \xi_1^{i_1}\cdots \xi_n^{i_n} 
\end{align}
To compute the above multiplication, we need to
move $\xi_k$ to site $k$ in a component of $\mathcal{A}$ by iteratively
swapping it with all $\xi_p$ in $\mathcal{A}$ with $p<k$. Since
$\xi_k$ is of odd parity, its swapping with $\xi_p^{i_p}$ would results in a
minus sign prefactor if $\xi_p^{i_p}$ is also of odd parity ($i_p=1$), i.e., 
\begin{align}
  \xi_k \xi_p^{i_p} = (-1)^{i_p} \xi_p^{i_p} \xi_k.
  \label{eq:swap_xi}
\end{align}
Thus we have
\begin{equation}
  \begin{split}
    \xi_k\mathcal{A}^{i_1\cdots i_n}=&[(-1)^{i_1}A^{i_1}\xi_1^{i_1}]\cdots [(-1)^{i_{k-1}}A^{i_{k-1}}\xi_{k-1}^{i_{k-1}}]\\
    &\times (A^{i_k}\xi_k\xi_k^{i_k})(A^{i_{k+1}}\xi_{k+1}^{i_{k+1}})\cdots(A^{i_n}\xi^{i_n}_n).
\end{split}
\end{equation}
Since $\xi_k\xi_k^{i_k=0}=\xi_k$ and $\xi_k\xi_k^{i_k=1}=0$, we may consider $\xi_k$ as a fermionic raising operator at site $k$: it creates a G-variable if the site is empty and sets it to zero if there is already a G-variable. Therefore, the total effect of the multiplication of $\xi_k$ can be represented as applying a matrix product operator (MPO) to an MPS $A^{i_1}\cdots A^{i_n}$:
\begin{equation}
  \label{eq:smpo-1}
  \includegraphics[]{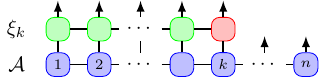}.
\end{equation}
This MPO consists of a string of sign operators (green node) before site $k$ and a raising operator (red node) at site $k$ and hence we call it \textit{signed matrix product operator} (SMPO). This string of signs is identical to that given by the Jordan-Wigner transformation \cite{JordanWigner1928}. Hence, by applying this SMPO to $A^{i_1}\cdots A^{i_n}$, we can obtain the desired G-tensor $\mathcal{B}$.

Now let us consider multiplying a quadratic term $\xi_k\xi_l$ (assuming $k<l$) to $\mathcal{A}$. In this case, we first move $\xi_l$
to site $l$. Such a move will apply sign operator on all sites $p<l$ and apply a raising operator on site $l$. Then we move $\xi_k$ to site $k$, which would affect all sites $q\le k$. The sign operator is applied twice on sites before $k$ and is thus canceled. Consequently, the corresponding SMPO can be written as
\begin{equation}
  \label{eq:smpo-2}
  \includegraphics[]{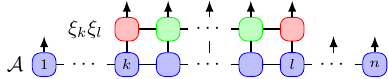}.
\end{equation}
The SMPO for more G-variables can be generalized accordingly.

If we multiply an exponent of G-variables, for instance
$e^{\xi_k\xi_j}$, to $\mathcal{A}$, then we have
\begin{equation}
  \label{eq:quadratic}
  e^{\xi_k\xi_j}\mathcal{A}=(1+\xi_k\xi_j)\mathcal{A}=\mathcal{A}+\mathcal{B}.
\end{equation}
This indicates that multiplying an exponent of G-variables to a G-tensor is equivalent to first applying an SMPO to get a new G-tensor, and then performing a summation of them. It should be noted that applying an SMPO, as defined in Eqs. \eqref{eq:smpo-1} and \eqref{eq:smpo-2} does not increase the bond dimension of the MPS. The bond dimension growth will be solely due to the MPS addition during this process. For simplicity, we also call the operation in form of \eqref{eq:quadratic} as an SMPO application.

\subsection{Multiplication of G-tensors}

We now introduce the multiplication between two G-tensors. Suppose we have two G-tensors
$\mathcal{A}=\mathcal{A}^{i_1\cdots
  i_n},\mathcal{B}=\mathcal{B}^{j_1\cdots j_n}$ and their
multiplication gives a new G-tensor $\mathcal{C}=\mathcal{A}\mathcal{B}$ in the form of 
\begin{equation}
\mathcal{C}^{i_1\cdots i_n}=C^{i_1\cdots i_n}\xi^{i_1}_1\cdots\xi_n^{i_n}.
\end{equation}
The direct product of $\mathcal{A}\mathcal{B}$ gives
\begin{equation}
  \label{eq:concat-AB}
   \mathcal{A}\mathcal{B}= A^{i_1}\cdots A^{i_n} \xi^{i_1}_1 \cdots \xi^{i_n}_n B^{j_1}\cdots B^{j_n} \xi^{j_1}_1 \cdots \xi^{j_n}_n. 
\end{equation}
To obtain $\mathcal{C}^{i_1\cdots i_n}$, we need to move all $\xi_k^{i_k}$ in $\mathcal{A}$ to the corresponding sites in $\mathcal{B}$, which gives an ``intermediate G-tensor'' with components
\begin{equation}
  \label{eq:intermediate-g-tensor}
  \mathcal{\tilde{C}}^{i_1j_1\cdots i_nj_n}=\tilde{C}^{i_1j_1\cdots i_nj_n}(\xi_1^{i_1}\xi_1^{j_1})\cdots(\xi_n^{i_n}\xi^{j_n}).
\end{equation}
Assuming that $\tilde{C}^{i_1j_1\cdots i_nj_n}$ is known, we can merge the index pairs $i_k,j_k$ into a single index with the following Grassmann multiplication relations 
\begin{equation}
 \begin{split}
  &\xi_k^{i_k=0}\xi_k^{j_k=0}=1,\quad\xi_k^{i_k=1}\xi_k^{j_k=1}=0,\\
  &\xi_k^{i_k=1}\xi_k^{j_k=0}=\xi_k^{i_k=0}\xi_k^{j_k=1}=\xi_k.
  \label{eq:grassmann-multiplication}
 \end{split}
\end{equation}
Therefore, the coefficient tensor
$C$ can be obtained from $\tilde{C}$ as
\begin{equation}
  \begin{split}
    &C^{i_k=0}=\tilde{C}^{i_k=0,j_k=0},\\
    &C^{i_k=1}=\tilde{C}^{i_k=1,j_k=0}+\tilde{C}^{i_k=0,j_k=1}.
  \end{split}
\end{equation}
% After this multiplication process, the desire $\mathcal{C}$ is obtained.

Now we need to find $\tilde{C}^{i_1j_1\cdots i_nj_n}$ from $A^{i_1\cdots i_n},B^{j_1\cdots j_n}$. In principle, we can first concatenate $\mathcal{A}$ and $\mathcal{B}$ as shown in Eq. \eqref{eq:concat-AB}, and then move the all $\xi_k$ to the corresponding positions in Eq. \eqref{eq:intermediate-g-tensor} by swapping G-variables. During this process, each swap is associated with a sign depending on the corresponding parity as specified by
\begin{equation}
  \xi_{p}^{i_{p}}\xi^{i_{q}}_{q}=(-1)^{i_{p}i_{q}}\xi^{i_{q}}_{q}\xi_{p}^{i_{p}},
\end{equation}
which means that swapping $\xi_p^{i_p}$ and $\xi_q^{i_q}$ would yield a minus sign when $i_p=i_q=1$. 

% e.g., swapping $\xi_p^{i_p}$ and $\xi_q^{i_q}$ would yield a minus sign when $i_p=i_q=1$. 

However, the total number of swaps required is huge, which could be very inefficient for MPS. To circumvent this issue, we can introduce auxiliary G-variables and represent   
\begin{equation}
\xi_1^{i_1}\cdots\xi_n^{i_n}=\int[\dd{\bm{\eta}}e^{\bm{\eta}}](\eta_1\xi_1)^{i_1}\cdots(\eta_n\xi_n)^{i_n},
\end{equation}
where $\int[\dd{\bm{\eta}}e^{\bm{\eta}}]=\int\dd{\eta_1}e^{\eta_1}\cdots\int\dd{\eta_n}e^{\eta_n}$ with $\eta$'s being the auxiliary G-variables. Since $(\eta_k\xi_k)^{i_k}$ contains two G-variables, it is of even parity and thus moving it does not result in extra signs. In this case, we can perform a direct tensor product of $A^{i_1}\cdots A^{i_n}$ and $B^{j_1}\cdots B^{j_n}$ site by site to obtain
\begin{equation}
  \begin{split}
    \tilde{\mathcal{C}}^{i_1j_1\cdots i_nj_n}&=(A_1\otimes B_1)^{i_1j_1}\cdots(A_n\otimes B_n)^{i_nj_n}\\
    &\times\int[\dd{\eta}e^{\eta}][(\eta_1\xi_1)^{i_1}\xi_1^{j_1}]\cdots
    [(\eta_n\xi_n)^{i_n}\xi_n^{j_n}].
  \end{split}
\end{equation}
All the auxiliary $\eta$'s can be moved to the leftmost by swapping, and then integrated out. At first glance, such a protocol still seems to require a huge amount of swapping operations, which would be as inefficient as before. However, these auxiliary G-variables can be fused and moved as a whole, making the process much more efficient than brute force swapping.

To do this, we first note that the following identity 
\begin{equation}
  \label{eq:eta-integral}
  \int[\dd{\bm{\eta}}e^{\bm{\eta}}]\eta_n^{i_n}\cdots\eta_1^{i_1}=1
\end{equation}
holds for any $i_1,\ldots,i_n=\qty{0,1}$, and it is the precondition of the fusion operation. Now let us move the rightmost $\eta_n$ to the
neighboring site of $\eta_{n-1}$. This gives
\begin{equation}
  \begin{split}
    &[(\eta_{n-1}\xi_{n-1})^{i_{n-1}}\xi_{n-1}^{j_{n-1}}][(\eta_n\xi_n)^{i_n}\xi_n^{j_n}]\\
    =&(-1)^{i_nj_{n-1}}[(\eta_n^{i_n}\eta_{n-1}^{i_{n-1}}\xi_{n-1}^{i_{n-1}})\xi_{n-1}^{j_{n-1}}]
       [\xi_n^{i_n}\xi_n^{j_n}],
  \end{split}
\end{equation}
where the sign is due to the movement of the  G-variable $\eta_n$ to the leftmost side in this expression. Due to the identity given by Eq. \eqref{eq:eta-integral}, the term $\eta_n^{i_n}\eta_{n-1}^{i_{n-1}}$ would eventually be integrated out to just scalar 1 anyway. Therefore, there is no need to keep track of all the details with respect to the indices $i_n,i_{n-1}$. Instead, keeping track of the parity information of these two G-variables as a whole is sufficient. Suppose these two G-variables are fused into a G-variable $\tilde{\eta}_{n-1}$, then we would have the odd and even parity parts as
\begin{equation}
  \begin{split}
  &\tilde{\eta}_{n-1}^{i_{n-1}=1}=\eta_n^{i_n=0}\eta_{n-1}^{i_{n-1}=1}+\eta_n^{i_n=1}\eta_{n-1}^{i_{n-1}=0},\\
   &\tilde{\eta}_{n-1}^{i_{n-1}=0}=\eta_n^{i_n=1}\eta_{n-1}^{i_{n-1}=1}+\eta_n^{i_n=0}\eta_{n-1}^{i_{n-1}=0}.
  \end{split}
\end{equation}
We can then move this newly fused G-variable to the neighbor of $\eta_{n-2}$, and repeat the fusion operation until all the auxiliary G-variables are moved to the leftmost end. Then the multiplication between $\xi$'s can be applied. Throughout this process, since we have only moved a site from the rightmost end to the leftmost end, the number of swapping operations required is much less than that needed for brute force swapping.

\section{Grassmann Matrix Product State}

The G-tensor multiplication described above reduced the number of swapping operations significantly. However, we can do better and avoid the swapping operation completely. To achieve this, we introduce the Grassmann matrix product state (GMPS). GMPS is a data structure to represent G-tensor $\mathcal{A}$ with the help of auxiliary G-variables. In GMPS, the components of G-tensor $\mathcal{A}$ is represented as
\begin{equation}
  \label{eq:gmps}
  \begin{split}
    \mathcal{A}^{i_1\cdots i_n}&=
    \sum_{\bm{\alpha},\bar{\bm{\alpha}}} A^{i_1}_{\alpha_1}\cdots A^{i_k}_{\bar{\alpha}_{k-1},\alpha_k}\cdots A^{i_n}_{\bar{\alpha}_{n-1}}\times\\
    &\int\mathcal{D}[\bar{\bm{\eta}}\bm{\eta}]\xi_1^{i_1}\eta_1^{\alpha_1}
      \cdots\bar{\eta}_{k-1}^{\bar{\alpha}_{k-1}}\xi_k\eta_k^{\alpha_k}\cdots\eta_{n-1}^{\bar{\alpha}_{n-1}}\xi_n^{i_n},
  \end{split}
\end{equation}
where the Grassmann space is enlarged by inserting the auxiliary G-variables $\bar{\eta}_k,\eta_k$ between $\xi_k$ and $\xi_{k+1}$. In our context, $\gK$ and $\gI$ are always of even parity, i.e., they contain an even number of G-variables. Therefore, we can require every site $A^{i_k}_{\bar{\alpha}_{k-1},\alpha_k}$ in Eq.~\eqref{eq:gmps} to satisfy the even parity condition
\begin{equation}
  \mod(\bar{\alpha}_{k-1}+i_k+\alpha_k,2)=0,
\end{equation}
where we assume $\bar{\alpha}_0=\alpha_n=0$ for the boundary sites. With such a condition, every $\eta_{k-1}^{\bar{\alpha}_{k-1}}\xi_k\eta_k^{\alpha_k}$ is of even parity, and thus they can be freely moved around as a whole without any sign issue. This allows us to multiply two GMPSs site by site easily.  Suppose we have two GMPSs $\mathcal{A}$ and $\mathcal{B}$ with site
tensors
\begin{equation}
  A^{i_k}_{\bar{\alpha}_{k-1}\alpha_k}\bar{\eta}_{k-1}^{\bar{\alpha}_{k-1}}\xi_k^{i_k}\eta_k^{\alpha_k},\quad
  B^{j_k}_{\bar{\beta}_{k-1}\beta_k}\bar{\eta}'^{\bar{\beta}_{k-1}}_{k-1}\xi_k^{j_k}\eta'^{\beta_k}_k.
\end{equation}
Multiplying these two site tensors gives 
\begin{equation}
  \begin{split}
  &A^{i_k}_{\bar{\alpha}_{k-1}\alpha_k}\otimes B^{j_k}_{\bar{\beta}_{k-1}\beta_k}
  (\bar{\eta}_{k-1}^{\bar{\alpha}_{k-1}}\xi_k^{i_k}\eta_k^{\alpha_k})
    (\bar{\eta}'^{\bar{\beta}_{k-1}}_{k-1}\xi_k^{j_k}\eta'^{\beta_k}_k)\\
    =&(-1)^{j_k\alpha_k}A^{i_k}_{\bar{\alpha}_{k-1}\alpha_k}\otimes B^{j_k}_{\bar{\beta}_{k-1}\beta_k}
       \bar{\eta}'^{\bar{\beta}_{k-1}}_{k-1}\bar{\eta}_{k-1}^{\bar{\alpha}_{k-1}}
       \xi_k^{i_k}\xi_k^{j_k}\eta_k^{\alpha_k}\eta'^{\beta_k}_k.
  \end{split}
\end{equation}
The auxiliary G-variables $\eta,\eta'$ can be fused together as discussed in the previous section, and the G-variables $\xi$ can be merged by the Grassmann multiplication defined in Eq.~\eqref{eq:grassmann-multiplication}. The site tensor form in Eq.~\eqref{eq:gmps} is thus restored and hence the result of multiplication of two GMPSs is still a GMPS.
\begin{figure}[htbp]
  \centerline{\includegraphics[]{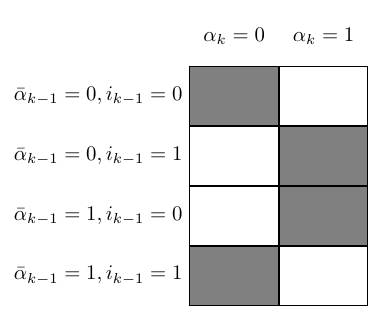}}
  \caption{Storage of a rank-3 site tensor. Only the gray blocks,
    which satisfy the even parity condition, are nonzero.}
  \label{fig:store}
\end{figure}

In the rank-3 site tensor $A^{i_k}_{\bar{\alpha}_{k-1}\alpha_k}$, only those components satisfying the even parity condition can be nonzero. In total, there are eight possible combinations for the three indices $\bar{\alpha}_{k-1},i_k,\alpha=\qty{0,1}$. Within these eight elements, half are nonzero. This property is shown in Fig. \ref{fig:store}, where the nonzero combinations are indicated by gray blocks. In numerical implementations, this allows us to save half of the memory usage.

In the following, by considering the single-orbital Anderson impurity model as an example, we shall demonstrate how to construct $\gK,\gI$ and evaluate the corresponding physical quantities using the GMS representation.

\section{Applications to single-orbital Anderson model}

The single-orbital Anderson model is a typical open quantum system problem, with its Hamiltonian given by 
$\Hop=\Hop_{\rm S}+\Hop_{\rm E}+\Hop_{\rm SE}$.  The impurity acts as the system that couples to the environment $\Hop_{\rm E}$ which we now refer to as the bath. The impurity Hamiltonian $\Hop_{\rm S}$ is 
\begin{equation}
  \Hop_{\rm S}=\varepsilon_d\sum_{\sigma}\adop_{\sigma}\aop_{\sigma}+
  U\adop_{\uparrow}\adop_{\downarrow}\aop_{\downarrow}\aop_{\uparrow},
\end{equation}
where $\varepsilon_d$ is the onsite energy for the impurity,
$\sigma={\uparrow,\downarrow}$ denotes the spin index, $U$ is the Coulomb interaction, and $\aop_{\uparrow,\downarrow}$ ($\adop_{\uparrow,\downarrow}$) are the annihilation (creation) operators of spin up and down fermion, respectively. The bath consists of free fermions $\hat{H}_{\mathrm{E}}=\sum_{\sigma k}\varepsilon_k\varepsilon_k\cdop_{\sigma k}\cop_{\sigma }$ and the impurity couples to the bath via the hybridization interaction 
\begin{equation}
  \hat{H}_{\mathrm{SE}}=\sum_{\sigma k}V_k(\adop_{\sigma}\cop_{\sigma k}+\cdop_{\sigma k}\aop_{\sigma}).
\end{equation}
The chemical potential of the bath is set to zero.

\subsection{Toulouse Model (Noninteracting $U=0$ Case)}

Now we demonstrate how to construct the GMPS of ADT $\gZ$ by starting from a simpler situation where the Coulomb interaction $U=0$. In this case, the electrons with different spin do not interact with each other and hence, we can omit the spin index in this section. The Anderson impurity model is thus reduced to an analytically solvable model, which is referred to as the Toulouse model \cite{leggett1987-dynamics} or the Fano-Anderson model \cite{mahan2000-many}.

\subsubsection{Construction of $\gK$}

To construct $\gK$, we follow the standard procedure given by Eqs. \eqref{eq:K}. After splitting $t_f=N\delta t$, the bare impurity propagator, together with the boundary G-variables, is expressed as
\begin{equation}
  \label{eq:Ka}
  \begin{split}
    \gK[\boldabar,\bolda]=&e^{-\abar a_N^+}e^{g \bar{a}_N^+a_{N-1}^+}\cdots e^{g \bar{a}_1^+a_0^+}\\
    &\times\rho_{\mathrm{S}}(\bar{a}_0^+,a_0^-)e^{\bar{g}\bar{a}_0^-a_1^-}\cdots e^{\bar{g}\bar{a}_{N-1}^-a_N^-}e^{\abar_N^-a},
  \end{split}
\end{equation}
where $g=e^{-i\delta t\varepsilon_d}$ and $\bar{g}$ is its complex conjugate. The initial condition is given by
\begin{equation}
\rho_{\mathrm{S}}(\abar_0^+,a_0^-)=\mel*{a_0^+}{\hat{\rho}_{\rm S}(0)}{a_0^-}.
\end{equation}

It can be seen that $\gK[\boldabar,\bolda]$ consists of exponents of time-local quadratic G-variables. Thus, the GMPS of $\gK[\boldabar,\bolda]$ can be constructed by sequentially applying these exponents to the vacuum state, which is represented by a GMPS with bond dimension one. Each application of the exponent can be implemented as an application of an SMPO as defined by Eq.~\eqref{eq:quadratic}.

For numerical evaluation, we need to specify an alignment of the G-variables to store the GMPS. It should be noted that the order of alignment of G-variables defined in the G-tensor Eq. \eqref{eq:g-tensor} can significantly affect the efficiency and the final storage performance of the construction. For instance, the most natural alignment of the G-tensor for $\gK[\boldabar,\bolda]$ may follow the direction of the Keldysh contour such that
\begin{equation}
  \abar a_N^+\bar{a}_N^+\cdots \bar{a}^+_0a^+_0a_0^-\bar{a}^-_0\cdots a_N^-\bar{a}_N^-a,
\end{equation}
where the boundary G-variables are placed at both ends. In this case, sequentially applying the exponents in Eq. \eqref{eq:Ka} can be illustrated as
\begin{equation}
  \label{eq:keldysh-alignment}
  \includegraphics[]{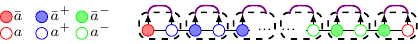},
\end{equation}
where the violet solid curve represents the application of the exponent of quadratic terms in the form of Eq. \eqref{eq:quadratic}. Each violet curve doubles the bond dimensions between the sites it applies to. Since the violet curves do not intersect with each other, the final G-tensor is built up with non-intersected blocks (wrapped up by a dashed box), whose inner bond dimension is two. This alignment fully reflects the time locality of the free impurity dynamics and corresponds to the direct product of their evolution operators.

Here we give another natural but inefficient alignment of G-tensors: we separate the G-variables by their conjugation and branch as
\begin{equation}
  a a_N^+\cdots a_0^+a_N^-\cdots a_0^-\bar{a}_0^-\cdots \bar{a}_N^-\bar{a}_0^+\cdots \bar{a}_N^+\abar.
\end{equation}
This alignment is natural in the sense that under G-integral, a G-tensor can be converted to an overlap of fermionic states, i.e., 
\begin{equation}
  \begin{split}
    &\int\mathcal{D}[\boldabar\bolda]\qty[a^i(a_N^+)^{i_N^+}\cdots (\abar_N^+)^{j_N^+}\abar^j]\\
    =&\mel{0}{\hat{a}^i(\hat{a}^+_N)^{i_N^+}\cdots \qty[(\hat{a}_N^+)^{\dag}]^{j_N^+}(\hat{a}^{\dag})^j}{0}.
\end{split}
\end{equation}
However, applying Eq.~$\eqref{eq:Ka}$ with such an alignment gives the following schematic figure
\begin{equation}
  \includegraphics[]{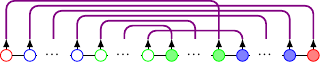}.
\end{equation}
It can be observed that there are many intersections between violet curves, where each intersection will double the bond dimensions it covers. The GMPS of $\gK$ constructed in this way is difficult to compress, i.e., the MPS compression can almost not reduce the bond dimension. Therefore, the bond dimensions would increase dramatically with this alignment.

Although the alignment defined by Eq. \eqref{eq:keldysh-alignment} is optimal for $\gK$, it is not friendly for $\gI$. A compromise should be made for global efficiency. We have thus chosen an alignment in which G-variables at the same time step are grouped together
\begin{equation}
  a \bar{a} a_N^{+}\bar{a}_N^+a_N^-\bar{a}_N^-\cdots a_0^+\bar{a}_0^{+}a_0^-\bar{a}_0^-.
\end{equation}
Such an alignment can be illustrated as
\begin{equation}
  \label{eq:toulouse-k-alignment}
  \includegraphics[]{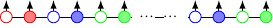}.
\end{equation}

\subsubsection{Construction of $\gI$}
\label{sec:gI}
The IF consists of several exponents of quadratic G-variables, thus the most direct way to construct its GMPS is by sequentially applying the corresponding SMPOs. The MPS compression algorithm is much more effective for $\gI$, but the alignment of the G-tensor would still greatly affects the bond dimensions. Here, we group G-variables at the same time step together, which gives the same alignment as \eqref{eq:toulouse-k-alignment} without the boundary G-variables.

Unlike the situation for $\gK$, the G-variables in exponents are correlated to each other in a time nonlocal manner. Because of this time nonlocality, the SMPO may cover a long distance and double the bond dimensions by its coverage, and we need to apply compression algorithm after every application of an SMPO. In total, we need $O(N^2)$ compression, which is inefficient. We seek a more efficient way following the spirit of TEMPO \cite{strathearn2018-efficient,strathearn2020-modelling,gribben2022-exact}. Instead of directly applying the SMPO, we write $\gI$ as a product of partial influence functionals (PIF) for which
\begin{equation}
  \label{eq:pif}
  \gI[\boldabar,\bolda]=\prod_{j,\zeta}\gI_j^{\zeta}[\abar_j\bolda]=\prod_{j,\zeta}e^{-\bar{a}_j^{\zeta}\sum_{k,\zeta'}\Delta^{\zeta\zeta'}_{jk}a^{\zeta'}_k}.
\end{equation}
Once the GMPS of the PIF $\gI_j^{\zeta}[\abar_j\bolda]$ is known, we can construct the IF by sequentially multiplication of PIFs. After every multiplication, we need one compression operation, giving only $O(N)$ compressions in total.

The PIF $\gI_j^{\zeta}[\abar_j\bolda]$ can be constructed by directly applying the SMPOs in Eq.~\eqref{eq:pif}. Constructing the PIF in this way requires $O(N)$ compressions, and constructing the IF still needs $O(N^2)$ compressions. At first glance, there is no improvement. However, since the bond dimension of the PIF is usually much smaller than that of IF, a much faster construction can be achieved. This method shares a similar idea as the iterative construction described in Ref. [\onlinecite{ng2023-real}].

Here is another method to construct the PIF. In a PIF $\gI_j^{\zeta}[\abar_j\bolda]$, there is a common G-variable $\bar{a}_j$ in all exponents. Thus, one can  expand the PIF as
\begin{equation}
  \label{eq:pif-expansion}
  \gI_j^{\zeta}[\abar_j\bolda]=1-\bar{a}_j^{\zeta}\sum_{k,\zeta'}\Delta^{\zeta\zeta'}_{jk}a^{\zeta'}_k,
\end{equation}
where only quadratic G-variable terms remain. Therefore, we need not take care the sign problem of higher-order G-variables. In this case, we may simply treat the G-variables as a raising operator acting on
the vacuum state. We can thus write the PIF as an operator
\begin{equation}
\gI_j^{\zeta}[\abar_j\bolda]=e^{-\hat{\bar{r}}_j^{\zeta}\sum_{k,\zeta'}\Delta^{\zeta\zeta'}_{jk}\hat{r}^{\zeta'}_k}
\end{equation}
acting on the history space. This operator can be represented as an MPO by the method discussed in Refs. [\onlinecite{gribben2022-exact,chen2024-gtempo}], and acting it on the vacuum state yields the desired PIF. Note that if the order of $\hat{\bar{r}}^{\zeta}_j$ and $\hat{r}_k^{\zeta'}$ needs to be swapped depending on the alignment of G-variables, then an extra minus sign will appear.

A third way which is more efficient and elegant to construct the PIF is to express Eq. \eqref{eq:pif-expansion} as a product of series matrices as
\begin{equation}
  \begin{split}
    \gI_j^{\zeta}[\abar_j\bolda]&=\mqty[1& \Delta^{\zeta\zeta'}_{j0}a^{\zeta'}_0]\cdots
                                           \mqty[1 & \Delta^{\zeta\zeta'}_{jj}a^{\zeta'}_j\\0 & 1]
                                                                                                \mqty[1&\abar^{\zeta}_j\\\abar^{\zeta}_j& 0]\\
       &\times\mqty[1&0\\-\Delta_{j,j+1}^{\zeta\zeta'}a^{\zeta'}_{j+1}&1]\cdots\mqty[1\\-\Delta_{jN}^{\zeta\zeta'}a_N^{\zeta'}].
  \end{split}
\end{equation}
With such an expression, and by observing that G-variables in PIF can be represented as a raising operator, we can directly write the PIF as an MPO with a bond dimension two \cite{guo2024-efficient}.

\subsubsection{Evaluation of Correlation Functions from ADT}

Once the GMPS of $\gK$ and $\gI$ are constructed, we can simply merge them as the ADT $\gZ[\boldabar,\bolda]=\gK[\boldabar,\bolda]\gI[\boldabar,\bolda]$. It should first be noted that the preferred alignment of the GMPS for $\gK$ and $\gI$ are not the same. The alignment \eqref{eq:toulouse-k-alignment} is preferred by $\gI$ but not the best alignment \eqref{eq:keldysh-alignment} for $\gK$. In practical calculations, since constructing $\gI$ is much more time-consuming, the alignment as illustrated in \eqref{eq:toulouse-k-alignment} is adopted at this time.

\begin{figure}[htbp]
\centerline{\includegraphics[]{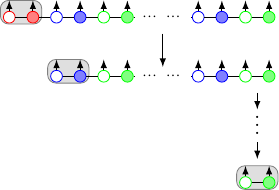}}
\caption{An illustration of the iterative G-integral of Eq. \eqref{eq:fermionic_corr}. The pairs of G-variables are integrated out iteratively according to Eq. \eqref{eq:g-integral}.}
\label{fig:g-integral}
\end{figure}

With the ADT, any multi-time correlation function (including equal-time observables) for the impurity can be evaluated via Eq. \eqref{eq:fermionic_corr}, where the G-integral over a pair of G-variables $\bar{\xi},\xi$ is evaluated by the relation
\begin{equation}
  \label{eq:g-integral}
  \int\dd{\bar{\xi}}\dd{\xi}e^{-\bar{\xi}\xi}A^{ij}\xi^i\bar{\xi}^j
  =A^{i=0,j=0}+A^{i=1,j=1}.
\end{equation}
The evaluation of Eq. \eqref{eq:fermionic_corr} can be achieved by iterative integration of the G-variables using the above formula. Such an iterative integration procedure is illustrated in Fig. \ref{fig:g-integral}.

\subsubsection{Zipup Algorithm}
\label{sec:zipup}

We now give a brief introduction of the zipup algorithm, which provides an efficient way to construct the ADT.  

A direct construction of the ADT $\gZ$ by multiplication of $\gK$ and $\gI$ would result in a large bond dimension $\chi_\gZ$ as $\chi_{\gZ} = \chi_{\gK}\chi_{\gI}$ where $\chi_{\gK}$ and $\chi_{\gI}$ are the bond dimensions of $\gK$ and $\gI$, respectively. The GMPS of $\gZ$ constructed in this way is difficult to compress and thus the memory cost to store the ADT is about $O(N\chi_{\gZ}^2)$. The corresponding computational cost of the G-integral calculation defined by Eq. \eqref{eq:fermionic_corr} is then about $O(N\chi_{\gZ}^3)$, which is unfriendly to numerical implementations when $\chi_{\gZ}$ is large.

This is not a problem for simple cases such as the Toulouse model, where the bond dimension of the ADT is $\chi_{\gK}\chi_{\gI}$, and $\chi_{\gK}$ is only 4 with alignment given by Eq. \eqref{eq:toulouse-k-alignment}. However, as we will discuss later, for the single-orbital Anderson impurity model, the bond dimension $\chi_{\gZ}$ becomes $\chi_{\gK}\chi_{\gI}^2$. Here the term $\chi_{\gI}^2$ is due to two spin flavors of the bath. Even with a modest $\chi_{\gI}\approx30$, this term yields $\chi_{\gZ}\propto\chi_{\gI}^2\approx1000$.

\begin{figure}[htbp]
\centerline{\includegraphics[]{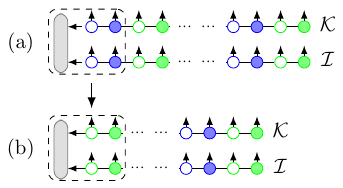}}
\caption{An illustration of the first step of zipup algorithm for Toulouse model.}
\label{fig:zipup}
\end{figure}

Hence, it is preferable to avoid the direct construction of $\gZ$ by multiplication of $\gK$ and $\gI$. Instead, we can construct them on the fly when computing expectation values given by Eq. \eqref{eq:fermionic_corr}. Such an algorithm is referred to as the zipup algorithm in Refs. [\onlinecite{chen2024-gtempo,chen2024-egtempo}], which boost the efficiency significantly. The basic idea of the zipup algorithm for Toulouse model is illustrated in Fig. \ref{fig:zipup}, where the boundary G-variables are omitted.

The algorithm starts with a trivial tensor $M$, which is denoted as the vertical gray bar in Fig. \ref{fig:zipup}(a). At the first step, this trivial tensor and the leftmost pair G-variables of $\gK$ and $\gI$ are multiplied together. The G-variables then are integrated out (indicated by the gray box) resulting in a new tensor [gray bar in Fig. \ref{fig:zipup}(b)]. The G-integral \eqref{eq:fermionic_corr} can be done via applying this procedure iteratively. The memory cost of the zipup algorithm is about $O(N(\chi_{\gK}^2+\chi_{\gI}^2))$ since we do not need to merge $\gK$ and $\gI$. This is clearly much less than $O(N\chi_{\gZ}^2)=O(N\chi_{\gK}^2\chi_{\gI}^2)$.

Let us now check the computational cost. In one step of the zipup algorithm, the tensors in the dashed box are multiplied together. The leg of the gray bar connecting $\gK$ has the bond dimension $\chi_{\gK}$, and the leg connecting $\gI$ has the bond dimension $\chi_{\gI}$. To multiply them together, we can first multiply $M$ and $\gK$, and this is equivalent to the multiplication of a $\chi_{\gK}\chi_{\gI}\times\chi_{\gK}$ matrix and a $\chi_{\gK}\times\chi_{\gK}$ matrix. The corresponding computational cost is about $O(\chi_{\gK}^2\chi_{\gI})$. Then we can multiply the $\gI$ part, the computational cost is then about $O(\chi_{\gK}\chi_{\gI}^2)$. In addition, during the zipup process, we need to swap the G-variables locally to the proper position for G-integral, which gives a cost of about $O(\chi_{\gK}^3+\chi_{\gI}^3)$. Therefore, the overall cost is about $O(N(\chi_{\gK}^3+\chi_{\gI}^3+\chi_{\gK}^2\chi_{\gI}+\chi_{\gK}\chi_{\gI}^2))$, which is significantly less than $O(N\chi_{\gZ}^3)=O(N\chi_{\gK}^3\chi_{\gI}^3)$.

\subsection{Single-orbital Anderson Impurity Model}
Now we proceed to the Anderson impurity model, where we need to consider the interaction term and G-variables with spin-up and spin-down flavors. The bare impurity propagator is then
\begin{equation}
  \label{eq:anderson-k}
  \begin{split}
    \gK[\boldabar,\bolda]=&\braket{-a}{a_N^+}\mel{a_N^+}{\hat{U}_{\rm S}}{a_{N-1}^+}\cdots\\
                          &\times\mel{a_0^+}{\hat{\rho}_{\rm S}(0)}{a_0^-}\cdots\\
    &\times\mel{a_{N-1}^-}{\hat{U}_{\rm S}^{\dag}}{a_N^-}\braket{a_N^-}{a},
  \end{split}
\end{equation}
where
$\ket*{a_k^{\pm}}=\ket*{a_{\uparrow k}^{\pm}a_{\downarrow k}^{\pm}}$ and 
\begin{align}
    &\mel{a_k^+}{\hat{U}_{\rm S}}{a_{k-1}^+}\nonumber \\
  = &
  e^{g\sum_{\sigma}\abar_{\sigma k}^+a_{\sigma,k-1}+g^2(e^{-i\delta tU}-1)
    \abar_{\uparrow k}^+\abar_{\downarrow k}^+a_{\downarrow,k-1}^+a_{\uparrow,k-1}^+},\\
  &\mel{a_k^+}{\hat{U}_{\rm S}^{\dag}}{a_{k-1}^+} \nonumber \\
  =&e^{\bar{g}\sum_{\sigma}\abar_{\sigma k}^+a_{\sigma,k-1}+\bar{g}^2(e^{i\delta tU}-1)
    \abar_{\uparrow k}^+\abar_{\downarrow k}^+a_{\downarrow,k-1}^+a_{\uparrow,k-1}^+},
\end{align}
with $g=e^{-i\varepsilon_d\delta t}$ and $\bar{g}$ being its complex conjugate. Once an alignment is chosen, we can apply the exponents in Eq.~\eqref{eq:anderson-k} to a vacuum GMPS sequentially to obtain the GMPS of $\gK$. In practice, we align these G-variables as (the boundary G-variables are omitted here)
\begin{equation}
  \begin{split}
  a_{\uparrow N}^+\abar_{\uparrow N}^+a_{\uparrow N}^-\abar_{\uparrow N}^-a_{\downarrow N}^+& \abar_{\downarrow N}^+  a_{\downarrow N}^-\abar_{\downarrow N}^-\cdots\\
  & a_{\uparrow 0}^+\abar_{\uparrow 0}^+a_{\uparrow 0}^-\abar_{\uparrow 0}^-a_{\downarrow 0}^+\abar_{\downarrow 0}^+a_{\downarrow 0}^-\abar_{\downarrow 0}^-.
  \end{split}
\end{equation}

\begin{figure}[htbp]
  \centerline{\includegraphics[]{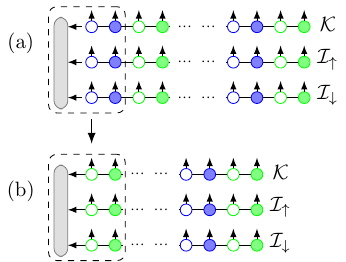}}
  \caption{An illustration of zipup algorithm for the Anderson impurity model.}
  \label{fig:zipup2}
\end{figure}

The IF for both spin flavors $\gI_{\uparrow}$ and $\gI_{\downarrow}$ are same as those of the Toulouse model, which can be obtained following the same procedure described in Sec. \ref{sec:gI}. Once the IF $\gI_{\uparrow/\downarrow}$ are known, we can use the zipup algorithm to calculate the desired quantities. The zipup algorithm is illustrated in Fig.~\ref{fig:zipup2}.

The zipup algorithm here deals with three GMPSs $\gK$ and $\gI_{\uparrow/\downarrow}$. Let $\chi_{\gI}$ be the bond dimension of $\gI_{\uparrow/\downarrow}$. It follows that the bond dimension of the ADT $\gZ$ will be $\chi_{\gK}\chi_{\gI}^2$ and we need $O(N\chi_{\gZ}^{2})\approx O(N\chi_{\gK}^2\chi_{\gI}^4)$ to store the ADT. The computational cost of the G-integral procedure for ADT would be about $O(N\chi_{\gZ}^3)\approx O(N\chi_{\gK}^3\chi_{\gI}^6)$. With the zipup algorithm, the memory cost would be reduced to $O(N(\chi_{\gK}^2+2\chi_{\gI}^2))$ and the computational cost would be reduced to $O(N(\chi_{\gK}^3+2\chi_{\gI}^3+\chi_{\gK}^{2}\chi_{\gI}^2+2\chi_{\gK}\chi_{\gI}^3))$.  The zipup algorithm can be generalized to handle more GMPSs straightforwardly, as discussed in Ref.~[\onlinecite{chen2024-egtempo}].
 
\section{Numerical examples}

In this section, we review some numerical results from Refs. [\onlinecite{chen2024-gtempo,chen2024-egtempo,chen2024-real,guo2024-efficient,GuoChen2024,ChenGuo2024c,guo2024-steady}] which demonstrate the capability of the GTEMPO method in various applications. These include real-time dynamics for nonequilibrium situations, real-time and imaginary-time equilibration dynamics, and spectral function extraction as an impurity solver. These results are benchmarked against other methods, including continuous-time quantum Monte Carlo, the tensor-network influence functional method \cite{ng2023-real}, and exact diagonalization.

\subsection{Real-time dynamics}

\begin{figure}
  \centering
  \includegraphics[width=\columnwidth]{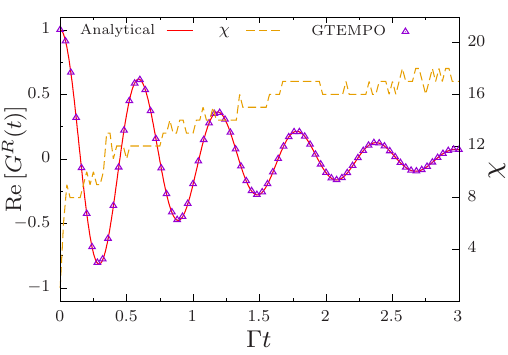}
  \caption{Figure from Ref. [\onlinecite{chen2024-gtempo}]. Benchmark of the retarded Green's function of the Toulouse model against the analytical solution. The yellow dashed line represents the bond dimension over time.}
  \label{fig:toulouse-retarded}
\end{figure}

To begin with, we show the benchmarks of GTEMPO results with the analytical solution for the Toulouse model. The comparison of the retarded Green's function between the GTEMPO results and the analytical solution are shown in Fig. \ref{fig:toulouse-retarded} modified from Ref. [\onlinecite{chen2024-gtempo}]. The GTEMPO results demonstrate a good agreement with the analytical solution. In addition, the growth of the bond dimension $\chi$ with respect to time is also shown by the yellow dashed line. Notably, the bond dimension almost stops growing at $\Gamma t\approx1.5$, where $\Gamma$ denotes the system-bath coupling strength and is used as the unit scale. The bond dimension required for the Toulouse model is around 16, which is very modest.

\begin{figure}[htbp]
  \centerline{\includegraphics[]{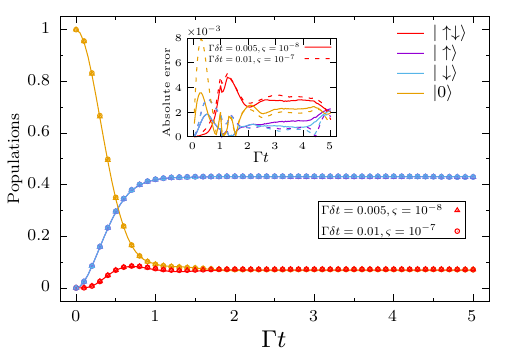}}
  \caption{Figure from Ref. [\onlinecite{chen2024-gtempo}]. Benchmark of the state populations against the results reported in Ref. [\onlinecite{ng2023-real}], which employs the tensor-network influence functional approach. }
  \label{fig:pop-vs-ng}
\end{figure}

In the case of Anderson impurity model where $U\ne 0$, we show the populations of different state in the system, which are of fundamental interests in Anderson impurity model. The transient dynamics of the population calculated by GTEMPO are benchmarked against other tensor-network influence functional approach in Ref. [\onlinecite{ng2023-real}], and the results are shown in Fig. \ref{fig:pop-vs-ng} \cite{chen2024-gtempo}. The system starts from the vacuum state $\ket{0}$, and after evolution it reaches the steady state where the populations of all four states remains nonzero values.

\begin{figure}
  \centering
  \includegraphics[width=\columnwidth]{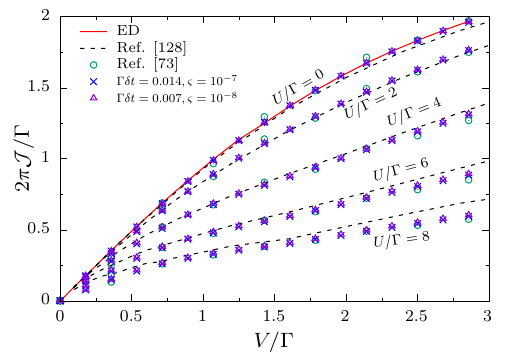}
  \caption{Figures from Ref. [\onlinecite{chen2024-gtempo}]. Current versus voltage for the two-bath single-orbital Anderson impurity model for different values of $U$, calculated by ED (red solid line for $U=0$), GTEMPO (purple triangles), Monte Carlo~\cite{bertrand2019-reconstructing} (black dashed lines), and tensor-network IF~\cite{thoenniss2023-efficient} (green circles). We have used the symmetrized particle current $\current = (\current_{\uparrow}^1 - \current_{\uparrow}^2) / 2 = (\current_{\downarrow}^1 - \current_{\downarrow}^2) / 2$.}
  \label{fig:current}
\end{figure}

Another useful application of GTEMPO is to study the nonequilibrium dynamics of the single-orbital Anderson impurity model with multiple baths. An advantage of the path integral formalism is that we just need to deal with a single hybridization function for all baths to compute the influence functional defined in Eq.~\eqref{eq:fermion_IF}, i.e.,
\begin{align}
  \Delta(\tau, \tau') = \sum_{\nu}\Delta^{\nu} (\tau, \tau'),
\end{align}
where $\Delta^{\nu}$ is the hybridization function for the $\nu$-th bath. As such, we need to construct only a single effective IF for all baths, rather than computing the IFs for different baths separately. 

We consider the nonequilibrium single-orbital Anderson impurity model where the impurity is coupled to two baths kept at different voltages. In such a scenario, the quantity of interest is the current, which is a key indicator of transport properties. We examine the particle current with spin $\sigma$ flowing out of the $\nu$-th bath, defined as the rate of change of the particle number $\langle \hat{N}^\mu_\sigma(t) \rangle$. It can be expressed by the system Green's function and the hybridization function as 
\begin{align}
  \current_{\sigma}^{\nu}(t) = -2 \Re \int_{\mathcal{C}} d\tau \Delta^{\nu}(t^+, \tau) \langle \bar{a}_{\sigma}(t^+) a_\sigma(\tau) \rangle
\end{align}
As shown in Fig. \ref{fig:current}, the currents calculated by GTEMPO are benchmarked with the quantum Monte Carlo method \cite{bertrand2019-reconstructing} and the tensor-network IF method~\cite{thoenniss2023-efficient}. The results for all three different methods show good consistency across almost all parameter regimes. This demonstrates the capability of GTEMPO in studying the nonequilibrium dynamics.

\subsection{Imaginary-time evolution for finite temperature equilibrium state}

The path integral can also be formulated on the imaginary-time axis, where the GTEMPO method can be directly applied. On the imaginary-time axis, the system partition function is expressed by imaginary-time Grassmann trajectories.

\begin{figure}
  \centering
  \includegraphics[width=\columnwidth]{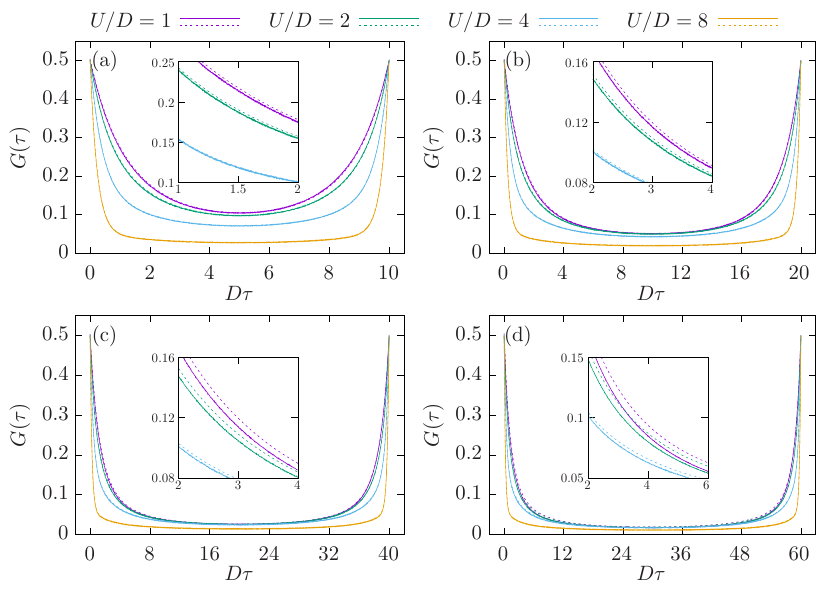}
  \caption{Figure from Ref. [\onlinecite{chen2024-egtempo}]. Matsubara Green’s function of the half-filling single-orbital AIM for (a) $\beta=10/D$, (b) $\beta=20/D$, (c) $\beta=40/D$, and (d) $\beta=60/D$. The dashed lines in each panel from top to bottom represent the GTEMPO results for $U/D=1,2,4,8$, and the corresponding CTQMC results are shown by the solid lines. Here, $D$ is the half bandwidth of the bath used as the unit scale. The inset is a zoom of the imaginary-time interval to better illustrate the differences among these results.}
  \label{fig:imaginary}
\end{figure}

The imaginary-time calculation aims at the equilibrium properties of the model, and the Matsubara Green's function is of particular interest. In Ref. [\onlinecite{chen2024-egtempo}], the Matsubara Green's function for single and two-orbital Anderson impurity models are computed and  benchmarked against the continuous-time Monte Carlo method \cite{gull2011-continuous}. Here, we just present the results for single-orbital Anderson model in Fig. \ref{fig:imaginary}.

An interesting phenomenon to note is that for GTEMPO, the imaginary-time calculation is more challenging than the real-time one, i.e., the bond dimension required for the imaginary-time calculation is generally much larger. This is because the imaginary path integral formalism exhibits the cyclic translational invariance, which is not suitable to be represented by open boundary MPS. An exception is the zero-temperature situation which we shall discuss later.

\subsection{Real-time impurity solver}

\begin{figure}
  \centering
  \includegraphics[width=\columnwidth]{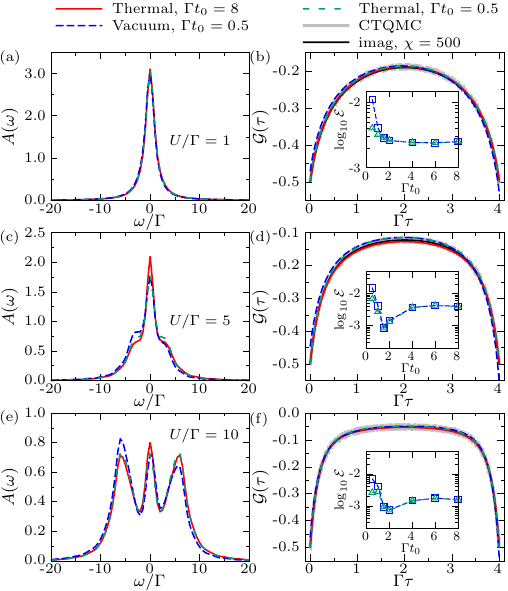}
  \caption{Figure from Ref. [\onlinecite{chen2024-real}]. (a,c,e) The spectral function $A(\omega)$ as a function of $\omega$ for different values of $U$. The red solid lines represent results calculated with $\Gamma t_0 = 8$ for the impurity thermal state. The blue and green dashed lines are results calculated with $\Gamma t_0 = 0.5$ for the impurity vacuum and thermal states, respectively. (b,d,f) The Matsubara Green's function $G(\tau)$ converted from the $A(\omega)$ calculated in (a,c,e) respectively. The black and gray solid lines in (b,d,f) represent the imaginary-time results for GTEMPO and CTQMC, respectively. The insets in (b,d,f) show the average errors of $G(\tau)$ between the real-time GTEMPO results (converted from the $A(\omega)$ calculated with different $t_0$) and the imaginary-time GTEMPO results. The blue dashed line with square and green dashed line with triangle are for the initial impurity vacuum and thermal states respectively. Here $\Gamma$ denotes the system-bath coupling strength parameter, which is used as the unit scale.}
  \label{fig:spectral}
\end{figure}

We can also solve the equilibrium Anderson impurity model on the real-time axis, where the GTEMPO method can be used as a real-time impurity solver \cite{chen2024-real}. The mainstream impurity solver is usually based on CTQMC, which is efficient only for imaginary-time calculations. Real-time information is extracted through analytical continuation, which is numerically ill-posed \cite{wolf2015-imaginary,fei2021-nevanlinna}, i.e., highly sensitive to noise. With GTEMPO, one can directly simulate the real-time dynamics of the impurity model. 

The basic procedures are summarized as follows: the model is prepared in the separate initial state Eq.~\eqref{eq:separate}, and then thermalized after some equilibration time $t_0$ by coupling to the bath. The retarded Green's function $G(t_0+t, t_0+t')$ is then calculated after this equilibration time. The spectral function $A(\omega)$ can be directly extracted from the retarded Green's function in frequency domain $G(\omega)$. This spectral function establishes a connection between the equilibrium retarded Green's function and the Matsubara Green's function, and can be used to calculate both the retarded and Matsubara Green's function.

In Fig. \ref{fig:spectral}, we show the spectral function $A(\omega)$ obtained by GTEMPO and the corresponding Matsubara Green's function obtained from $A(\omega)$. We benchmark these results against those from CTQMC for different $U$. For different initial states, the GTEMPO results represented by the dashed lines are in good agreement with the CTQMC results represented by the gray shades.

\section{Further developments}

The GTEMPO method can also be applied to the L-shaped Kadanoff-Baym contour \cite{kadanoff1962-quantum,aoki2014-nonequilibrium}, where both the imaginary-time and real-time information can be obtained simultaneously \cite{ChenGuo2024c}. The construction procedures of $\gK$ and $\gI$ are exactly the same as those for the partial influence functional method described in this article. The only difference is that we need to replace the path integral formalism on the Keldysh contour with that on the Kadanoff-Baym contour.

On the Keldysh contour, the hybridization function \eqref{eq:hybridization} is time-translationally invariant. This property has not been utilized in the partial influence functional method described in this article. The implementation of time-translational invariance allows the influence functional to be constructed more efficiently \cite{guo2024-efficient}. In such a situation, we only need to store and manipulate GMPS sites within a single time step, and the number of GMPS multiplication needed to construct $\gI$ is almost independent of the evolution time.

At zero temperature ($\beta=\infty$), the hybridization function differs from that at finite temperature as it always decays. Thus, the boundary condition information would eventually be lost, and open boundary MPS can again be used effectively. Moreover, at zero temperature, the hybridization also exhibits imaginary-time translational invariance. Therefore, we can also use the infinite MPS technique to construct $\gK$ and $\gI$ and directly access the ground-state information \cite{GuoChen2024}.

In most of the path-integral-based methods we have so far, including QuAPI, TEMPO, and GTEMPO, the steady state is obtained through finite-time evolution, which can still be a long transient dynamics. In fact, by utilizing time translational invariance and the infinite MPS technique, we can directly target the infinite time limit where the steady state is reached. Here, the steady state can be either an equilibrium or a nonequilibrium one. In such infinite time calculations, the transient dynamics information is completely lost, and only the GMPS sites within a single time step need to be stored and manipulated. At this time, the infinite GMPS is most recommended for both equilibrium and nonequilibrium steady state calculations, and more details can be found in Ref. [\onlinecite{GuoChen2024}].

\section{Conclusions}

We present a general overview of the Grassmann time-evolving matrix product operator method, a numerical implementation for fermionic influence functional to study open quantum systems. As its name suggests, it shares a similar spirit with the bosonic version TEMPO \cite{strathearn2018-efficient}. We first review the formalism for both the bosonic and the fermionic path integral to give a clear picture the difficulty in Grassmann path integral evaluations. We then dive into the details of Grassmann tensors, signed matrix product operators, and the Grassmann matrix product states, which are developed to tackle the Grassmann algebra when operating these influence functional tensors. The construction of these objects is demonstrated with the single-orbital Anderson impurity model. The numerical benchmarks are also demonstrated to show real-time nonequilibrium dynamics, real-time and imaginary-time equilibration dynamics, and solving the spectral functions for impurity models. Some further developments are introduced mainly focused on the infinite GMPS, which is main focus of this article. 

For future studies, we aim to incorporate the impurity solver in the DMFT calculations to study strongly-correlated many-body systems. Other potential applications include the study of strong-coupling quantum thermodynamics where a non-perturbative treatment is crucial. Moreover, we believe that the infrastructure of Grassmann tensor network can also be applied to other problems involved with Grassmann algebra. Lastly, we highlight that the open source package for GTEMPO introduced in this article will be available soon.

\section{Acknowledgement}
This work is supported by the NSFC Grant Nos. 12104328 and 12305049. C. G. is supported by the Open Research Fund from State Key Laboratory of High Performance Computing of China (Grant No. 202201-00).

% \bibliography{refs.bib, tempo.bib}

%aipnum4-2.bst 2019-01-14 (MD) hand-edited version of apsrev4-1.bst
    
\end{document}